# A Precise Information Flow Measure from Imprecise Probabilities

Sari Haj Hussein
Department of Computer Science, Aalborg University, Denmark
Email: angyjoe@gmail.com

*Abstract*—Dempster-Shafer theory of imprecise probabilities has proved useful to incorporate both nonspecificity and conflict uncertainties in an inference mechanism. The traditional Bayesian approach cannot differentiate between the two, and is unable to handle non-specific, ambiguous, and conflicting information without making strong assumptions. This paper presents a generalization of a recent Bayesian-based method of quantifying information flow in Dempster-Shafer theory. The generalization concretely enhances the original method removing all its weaknesses that are highlighted in this paper. In so many words, our generalized method can handle any number of secret inputs to a program, it enables the capturing of an attacker's beliefs in all kinds of sets (singleton or not), and it supports a new and precise quantitative information flow measure whose reported flow results are plausible in that they are bounded by the size of a program's secret input, and can be easily associated with the exhaustive search effort needed to uncover a program's secret information, unlike the results reported by the original metric.

*Index Terms*—computer security, quantitative information flow, imprecise probabilities, Dempster-Shafer theory, information theory, uncertainty, inference, program analysis

## I. Introduction

The goal of information flow analysis is to enforce limits on the use of information that apply to all computations that involve that information. For instance, a confidentiality property requires that a program with secret inputs should not leak those inputs into its public outputs. Qualitative information flow properties, such as non-interference are expensive, impossible, or rarely satisfied by real programs: generally some flow exists, and many systems remain secure provided that the amount of flow is sufficiently small, moreover, designers wish to distinguish acceptable from unacceptable flows.

Systems often reveal a summary of secret information they store. The summary contains fewer bits and provides a limit on the attacker's inference. For instance, a patient's report is released with the disease name covered by a black rectangle. However, it is not easy to precisely determine how much information exists in the summary. For instance, if the font size is uniform on the patient's report, the width of the black rectangle might determine the length of the disease name. Quantitative information flow (QIF) analysis is an approach that establishes bounds on information that is leaked by a program. In QIF, confidentiality properties are also expressed, but as limits on the number of bits that might be revealed from a program's execution. A violation is declared if the number of leaked bits exceeds the policy.

The metric in [1] is based on a new perspective for QIF analysis. The fundamental idea is to model an attacker's belief about a program's secret input as a probability distribution over high states. This belief is then revised, using Bayesian updating techniques, as the attacker interacts with a program's execution. It is believed that the work reported in [1] is the first to address an attacker's belief in quantifying information flow. This work was later expanded and appeared in [2]. A number of relevant results [3], [4] were reported in the sequel; however, the work in [1], [2] is sufficient as a foundation of our work.

A number of weaknesses can be seen in [2]. First, probability measures are used for capturing an attacker's belief and representing *her* uncertainty about the true state of a system. These measures have the *finite additivity property* that forces them to act on singleton sets, and makes it difficult to represent an attacker's ignorance or contradiction. Moreover, these measures *cannot* model attackers who effectually or ineffectually collaborate with each other. Second, the experiment protocol between an attacker and a system described in [2] *cannot* handle more than one secret input to a program. Third, the QIF metric advanced in [2] reports counter-intuitive flow quantities that *exceed* the size of a program's secret input, and make it *impossible* to determine the space of the exhaustive search needed to uncover a program's secret information.

This paper presents a generalization of the method followed in [2] that is free of all these weaknesses. The generalization is based on Dempster-Shafer theory of imprecise probabilities [5], [6] which enables the capturing of an attacker's beliefs in all kinds of sets (singleton or not), combining those beliefs, and revising them to update an attacker's knowledge about a system. As part of this generalization, we propose an inference scheme an attacker uses to update her knowledge from interacting with a program execution. This scheme can handle *any* number of secret inputs to a program. The mathematical toolbox on beliefs and the inference scheme we posit in this paper support a new and precise QIF measure whose reported flow results are *bounded* by the size of a program's secret input, and can be *easily* associated with the exhaustive search effort needed to uncover a program's secret information, unlike the results reported by the original metric.

### A. Relation to Our Earlier Work

In a recent position paper [7], we tackled the inexplicable results reported by the QIF metric in [2] that exceed the size of a program's secret input, and presented a *refinement* that bounds those results by a range consistent with the size of a program's secret input. The refinement was accomplished under the *original* Bayesian settings, and it enabled us to relate the reported flow results to the exhaustive search effort needed to uncover a program's secret information. A reader, interested in developing a clear picture of the problems the metric in [2] is fraught with, is strongly referred to [7].

*B. Plan of the Paper*

The remainder of this paper is organized as follows. Section II discusses the methods of representing uncertainty starting from the coarse-grained frame of discernment, moving to joint frames, tuples, and tuple sets, and ending with the fine-grained belief functions. In this section, we rigorously clarify the limitations of probability measures used in [2]. Section III concentrates on capturing beliefs using mass functions and the transformation of these functions into belief functions. Our mathematical toolbox on beliefs is given in Section IV. It includes formulas for combining beliefs, conditioning them, and measuring the divergence between them. In this section, we give a clear comparison between the poor properties of Kullback-Leibler divergence measure [8] (the authors' choice in [2]), and the appealing ones of Jensen-Shannon divergence measure [9] (our choice). We further investigate and *succeed* in generalizing Jensen-Shannon divergence measure in Dempster-Shafer theory. Section V presents the language needed in our experiments. Section VI lifts the syntax and semantics of this language in order to enable us to write programs source code in terms of mass functions. Section VII gives the attacker's model and then presents an inference scheme an attacker uses to update her knowledge from interacting with a program execution. Section VIII experiments with this inference scheme using various set structures induced by an attacker's beliefs. Our informal reasoning and generic observations about experiments' results are also given in this section. Section IX deals with quantifying information flow and advances a new and precise QIF measure whose reported flow results are proved to be bounded by the size of a program's secret input, and easily associated with the exhaustive search effort needed to uncover a program's secret information. Sample flow calculations are also given in this section. The paper concludes in Section X.

*C. Novel Contributions*

We believe that the work reported herein is the *first* to address the use of Dempster-Shafer theory in quantifying information flow. A number of novel contributions that, to the best of our knowledge do not appear in the literature, are also seen over the course of this correspondence. They are the *generalization* of Jensen-Shannon divergence measure in Dempster-Shafer theory, the *rules* of updating a mass function, and conditioning it on a Boolean expression, in addition to the lifted imperative while-language that *acts* on mass functions. All the uncertainty computations that appear in this paper are worked out using the *pyuds library* [10]; a Python library we developed specifically for this purpose.

## II. REPRESENTING UNCERTAINTY

*A. Frame of Discernment*

For most representations of uncertainty, the starting point is a set of possible *worlds*, *states*, or *elementary outcomes* that an agent considers possible. This set is called a *frame of discernment* [11] (a frame for short). For example, in the crude guessing of commonly used passwords, an agent might consider the following set possible:

$$\{\text{password}, 123456, \text{qwerty}, \text{abc123}, \text{letmein}, \text{monkey}, 696969\}$$

The frames dealt with in this paper are given under the *closed-world assumption* [5]. For a finite frame $\mathcal{W} = \{w_1, ..., w_n\}$, this means two things:

1) Exclusiveness: The worlds $w_i$ in $\mathcal{W}$ are mutually exclusive which means that *at most* one of them is the true world.
2) Exhaustiveness: The frame $\mathcal{W}$ is complete which means that it contains *all* the possible worlds.

A state in a program execution is an assignment of a value to a variable, and a frame is eligible to contain a set of those assignments. For instance, a Boolean variable $a$ accepts two possible assignments $a \to 0$ or $a \to 1$. It has two possible states that we may write as $\sigma = (a \to 0)$ and $\sigma = (a \to 1)$, and its corresponding frame is $\mathcal{W}_a = \{0, 1\}$.

*B. Joint Frame, Tuple, and Tuple Set*

A program execution may accept a number of *secret (high)* and *nonsecret (low)* inputs. For each input, we have a number of possible states that we should assimilate into an independent frame. To represent an agent's uncertainty about these two types of inputs, we need to define the notions of joint frame, tuple, and tuple set [12].

*Definition 1 (Joint Frame):* Let $r$ be a finite universal variable set where for each variable $X \in r$ there exists a frame $\mathcal{W}_X$ of values that can be assigned to $X$, and let $s \subseteq r$ be a variable set. The joint frame on $s$ is defined by the formula:

$$\mathcal{W}_s = \prod_{X \in s} \mathcal{W}_X$$

*Definition 2 (Tuple):* Let $\mathcal{W}_s$ be a joint frame on $s \subseteq r$. An $s$-tuple is a function of the form $x : s \to \mathcal{W}_s$ that associates a value $x(X) \in \mathcal{W}_s$ with each variable $X \in s$.

*Definition 3 (Tuple Set):* Let $\mathcal{W}_s$ be a joint frame on $s \subseteq r$. An $s$-tuple set is a subset $S \subseteq \mathcal{W}_s$.

Definitions 1-3 allow us to assume two joint frames; a *high* joint frame $\mathcal{W}_h$ on a *high* variable set $h \subseteq r$, and a *low* joint frame $\mathcal{W}_l$ on a *low* variable set $l \subseteq r$, to represent an agent's uncertainty about secret and nonsecret inputs respectively. The overall joint frame $\mathcal{W}_{h \cup l}$ on the overall variable set $h \cup l \subseteq r$ emerges as the product of these two frames:

$$\mathcal{W}_h = \prod_{X \in h} \mathcal{W}_X, \mathcal{W}_l = \prod_{X \in l} \mathcal{W}_X, \mathcal{W}_{h \cup l} = \prod_{X \in h \cup l} \mathcal{W}_X$$

In the remainder of this correspondence, a frame is always joint unless we state otherwise. When we refer to a frame, we write $\mathcal{W}_s$, however we do not say that it is taken on the variable set $s \subseteq r$. In addition, states are handled similarly to tuples, and likewise state sets to tuple sets. When we say the high and low projections of a state, we mean the projections of that state to $h$ and $l$ respectively.

## C. Belief Functions

A frame is a coarse-grained representation of uncertainty, since we do not have any means of *comparing* the likelihood of two worlds. Belief functions, the cornerstones of Dempster-Shafer theory [5], [6], offer a fine-grained representation of uncertainty that is suitable for our work because they are *numeric* thus enabling us to *quantitatively* measure information flow. They further permit the modeling of the *evolution* (or *regression*) of an agent's knowledge about a system as more and more pieces of evidence become available. Additionally, they admit a programming language semantics, as we will show in Section VI. Finally, under belief functions, all pairs of worlds are *comparable* thus promoting the reasoning of agents and empowering our analysis.

Although probability measures, the authors' choice in [2], are familiar, quantitative, support operations on beliefs, and admit a programming language semantics, they have the *finite additivity property* that forces them to act on singleton sets. This makes it difficult to represent *ignorance* (by assigning a zero probability to a set in an algebra) and *contradiction* (by assigning a nonzero probability to the empty set). It also complicates assigning probabilities to *non-singleton* and *joint* sets. The inability of agents to capture ignorance, express contradiction, and believe in non-singleton and joint sets clearly *detracts* from the depth of our analysis. In addition, probability measures entail assigning scalar probabilities to *all* sets in an algebra, but an agent may *not* have sufficient computational power to do that. This computational inefficiency escalates into a grueling ordeal when dealing with huge frames. Lastly, probability measures can *only* capture independent work, while *failing* at modeling attackers who effectually or ineffectually collaborate with each other as rigorously clarified in Example 1.

*Example 1 (Modeling Attackers' Collaboration):* Consider a band of attackers whose purpose is to hack into a computer system. Assume that this band is partitioned into sub-bands $A_1$, $A_2$,..., $A_n$ and let $\mu(A_i)$ be the degree of infiltration begotten by the sub-band $A_i$. For any two sub-bands $A_i$ and $A_j$, it is intuitive that any of the following can happen:

- $\mu(A_i \cup A_j) = \mu(A_i) + \mu(A_j)$ when $A_i$ and $A_j$ work independently.
- $\mu(A_i \cup A_j) > \mu(A_i) + \mu(A_j)$ when $A_i$ and $A_j$ effectually collaborate.
- $\mu(A_i \cup A_j) < \mu(A_i) + \mu(A_j)$ when $A_i$ and $A_j$ ineffectually collaborate.

## III. CAPTURING BELIEF

A belief is a *psychological state* in which an agent has a degree of support to a proposition about a system. A belief is based on a piece of evidence an agent obtains through some mean. In the framework of Dempster-Shafer theory, this belief is captured using a mass function, which is defined as follows.

*Definition 4 (Mass Function):* Let $\mathcal{W}_s$ be a frame. A mass function on $\mathcal{W}_s$ is a function of the form $m : \mathcal{P}(\mathcal{W}_s) \to [0, 1]$ where $\mathcal{P}(\mathcal{W}_s)$ is the first-order power set of $\mathcal{W}_s$ defined as $\mathcal{P}(\mathcal{W}_s) = \{X | X \subseteq \mathcal{W}_s\}$. This function satisfies:

$$m(\emptyset) = 0, \sum_{A \in \mathcal{P}(\mathcal{W}_s)} m(A) = 1$$

For any $A \in \mathcal{P}(\mathcal{W}_s)$, the value $m(A)$ has the following meaning; it characterizes the degree of belief that the true world is in the tuple set $A$, but it does *not* take into account any additional evidence for the various subsets of $A$.

Each tuple set $X \in \mathcal{P}(\mathcal{W}_s)$ such that $m(X) > 0$ is called a *focal set* of $m$. We denote the set of all focal sets induced by $m$ as $\mathcal{F}_m$, and write:

$$\mathcal{F}_m = \{X \in \mathcal{P}(\mathcal{W}_s) | m(X) > 0\}$$

We call the pair $\langle \mathcal{F}_m, m \rangle$ a *body of evidence*. Occasionally, we denote the domain $\mathcal{P}(\mathcal{W}_s)$ of $m$ as $d(m)$. Definition 5 shows how to project a mass function.

*Definition 5 (Mass Function Projection):* Let $\mathcal{W}_s$ be a frame, $m : \mathcal{P}(\mathcal{W}_s) \to [0, 1]$ be a mass function on $\mathcal{W}_s$, and $t \subseteq s$ be a variable set. The projection of $m$ to $t$ is defined for any $A \in \mathcal{P}(\mathcal{W}_t)$ by the formula:

$$m^{\downarrow t}(A) = \sum_{B^{\downarrow t} = A} m(B)$$

where $B^{\downarrow t}$ is the projection of the tuple set $B \in \mathcal{P}(\mathcal{W}_s)$ to $t$.

As a specialization of the general mass function, we define a point mass function as follows.

*Definition 6 (Point Mass Function):* Let $\mathcal{W}_s$ be a frame, and $m : \mathcal{P}(\mathcal{W}_s) \to [0, 1]$ be a mass function on $\mathcal{W}_s$. We say that $m$ is a point on the tuple set $A \in \mathcal{P}(\mathcal{W}_s)$, and write $\dot{m}_A$, if the degree of belief characterized by $m$ is *fully concentrated* on $A$, that is, if $m(A) = 1$.

Since it does *not* have the finite additivity property, a mass function $m$ is *not* a measure. This can be coped with. One can *bind* the pieces of evidence together, and obtain a belief measure from $m$ using the formula:

$$Bel(A) = \sum_{B \subseteq A} m(B)$$

Since the tuple sets in the domain of the function $Bel : \mathcal{P}(\mathcal{W}_s) \to [0, 1]$ are measurable, normalizing the values $Bel(A)$, so that the sum is 1, allows us to apply the familiar distribution arithmetic on them i.e., distribution sum, product, conditioning, and difference [13]. However, this is *not* what we want to do. Converting the values $m(A)$ to $Bel(A)$ is an *expensive* operation that should be kept to a minimum. Moreover, dealing with the values $m(A)$ is more tractable than dealing with $Bel(A)$. Thus, we ought to maintain the mass function setting in our work and propose the following arithmetic on beliefs.

## IV. ARITHMETIC ON BELIEFS

### A. Belief Combination

We combine beliefs using Dempster's combination rule [14]. Given two pieces of evidence obtained from two *independent* sources (we will shortly discuss independence) and expressed by two mass functions $m_1$ and $m_2$ on the *same* frame $\mathcal{W}_s$, Dempster's combination rule aggregates $m_1$ and $m_2$ to obtain a combined mass function $m_1 \otimes m_2$ which is defined for any tuple set $\emptyset \neq A \in \mathcal{P}(\mathcal{W}_{s \cup s})$ by the formula:

$$(m_1 \otimes m_2)(A) = k. \sum_{B \cap C = A} m_1(B).m_2(C) \quad (1)$$

where:

$$(m_1 \otimes m_2)(\emptyset) = 0, k^{-1} = \sum_{B \cap C \neq \emptyset} m_1(B).m_2(C)$$

If $m_1$ and $m_2$ are defined on two different frames $\mathcal{W}_s$ and $\mathcal{W}_t$, then the intersection $B \cap C$ is inapplicable anymore and is replaced with the *natural join operation* $B \bowtie C$ [12] as expressed by the formula, which is defined for any tuple set $\emptyset \neq A \in \mathcal{P}(\mathcal{W}_{s \cup t})$:

$$(m_1 \otimes m_2)(A) = k. \sum_{B \bowtie C = A} m_1(B).m_2(C) \quad (2)$$

where:

$$(m_1 \otimes m_2)(\emptyset) = 0, k^{-1} = \sum_{B \bowtie C \neq \emptyset} m_1(B).m_2(C)$$

The parameter $k$ in formulas (1) and (2) normalizes $m_1 \otimes m_2$ which has the appeal of explicitly recognizing *conflict* between the pieces of evidence an agent gathers about a system [15].

A prerequisite for using Dempster's combination rule is that the pieces of evidence are obtained from independent sources. Intuitively, this means that these pieces are totally *unrelated* and that the occurrence of one of them has no influence on the other [11]. In our work, this is well-justified if the pieces of evidence are obtained from external sources that are unrelated to a program execution; however, it is *not* if the pieces are obtained by monitoring an execution - in repeated executions, an agent relies on one output to rearrange the next input and thus influence the next output [2].

Dempster's combination rule has the distinguishing property of being commutative and associative [11]. This empowers our analysis by allowing an agent to choose the *combination* order and postpone the combination of a *misleading* piece of evidence until more hints about this piece are available.

### B. Belief Conditioning

We condition beliefs using Dempster's conditioning rule [14]. Suppose that a current agent's belief is captured using a mass function $m : \mathcal{P}(\mathcal{W}_s) \rightarrow [0,1]$. Later on, this agent obtains a new piece of evidence that the true world is in the tuple set $B \in \mathcal{P}(\mathcal{W}_s)$. Suppose further that there exists a focal set $C \in \mathcal{F}_m$ such that $C \cap B \neq \emptyset$. Dempster's conditioning rule enables the agent to *incorporate* the new evidence and *update* her knowledge. This rule transforms $m$ into a new mass function $m_B$ as expressed by the formula, which is defined for any tuple set $\emptyset \neq A \in \mathcal{P}(\mathcal{W}_s)$:

$$m_B(A) = \begin{cases} k. \sum_{C \cap B = A} m(C) & \text{for } A \neq \emptyset \\ 0 & \text{for } A = \emptyset \end{cases} \quad (3)$$

where:

$$k^{-1} = \sum_{C \cap B \neq \emptyset} m(C)$$

The parameter $k$ has the effect of normalizing $m_B(A)$, and enjoys the same quality mentioned in the previous section.

### C. Belief Divergence

*1) Choosing a Divergence Measure:* An agent's belief about a program's secret input is modeled as a probability distribution in [2], and the divergence between two probability distributions is measured using Kullback-Leibler divergence [8], which is given in Definition 7.

*Definition 7 (Kullback-Leibler Divergence Measure):* Let $X$ be a discrete random variable with alphabet $\mathcal{X}$, and let $p_1$ and $p_2$ be two probability distribution functions on $X$. The Kullback-Leibler divergence measure between $p_1$ and $p_2$ is defined by the formula:

$$KL(p_1, p_2) = \sum_{x \in \mathcal{X}} p_1(x) \log \frac{p_1(x)}{p_2(x)}$$

Our work necessitates a divergence measure between mass functions, not between probability distributions. $KL$ divergence *cannot* be written in terms of generalizable uncertainty functionals, and thus *seems* non-generalizable in Dempster-Shafer theory to act on mass functions. In contrast, Jensen-Shannon divergence measure [9] has an obvious information-theoretic interpretation in terms of Shannon uncertainty functional, which makes it generalizable in Dempster-Shafer theory, in addition to a number of desirable properties that $KL$ lacks. Before defining Jensen-Shannon divergence measure, we need to give a definition for Shannon uncertainty functional.

*Definition 8 (Shannon Uncertainty Functional):* Let $X$ be a discrete random variable with alphabet $\mathcal{X}$, and let $p$ be a probability distribution function on $X$. The uncertainty about $X$ is defined by the functional:

$$S(p) = -\sum_{x \in \mathcal{X}} p(x) \log p(x)$$

Uncertainty is measured in bits if the logarithm is binary. (Here and hereafter, all logarithms are to the base 2).

*Definition 9 (Jensen-Shannon Divergence Measure):* Let $p_1$ and $p_2$ be two probability distribution functions. The Jensen-Shannon divergence measure between $p_1$ and $p_2$ is defined by the formula:

$$JS(p_1, p_2) = 2S(\frac{p_1 + p_2}{2}) - S(p_1) - S(p_2)$$

In Table I, we compare between $KL$ and $JS$ divergence measures. $P3$ is a salient property that maintains the *balance* and computational correctness in the information flow measure we will advance in Section IX. $P4$ is important in its own right, since it enables us to handle *all* possible belief combinations, including those where one belief is zero and the other is positive. The dissatisfaction of $P4$ in $KL$ drives the authors of [1] to suggest an admissibility restriction on beliefs whose *ineffectiveness* is revealed in our earlier work [7]. We also see that P6 is appealing to have in our work. Indeed, it decidedly contributes to the desirable boundness of the flow measure we will propose in Section IX.

TABLE I
COMPARISON BETWEEN $KL$ AND $JS$ DIVERGENCE MEASURES

| No | Property | $KL$ | $JS$ |
|---|---|---|---|
| P1 | $D(p_1, p_2) \geq 0$ iff $p_1(x) \neq p_2(x)$ | Yes | Yes |
| P2 | $D(p_1, p_2) = 0$ iff $p_1(x) = p_2(x)$ | Yes | Yes |
| P3 | $D(p_1, p_2) = D(p_2, p_1)$ | No | Yes |
| P4 | Finiteness (Definement) | Not if we have $p \log \frac{p}{0}$ | Yes |
| P5 | Upper and lower bounds | No, only lower bound | Yes |
| P6 | Boundness | No | Yes, $JS \leq 2$ |

*2) Generalizing the Divergence Measure:* As we saw in Definition 9, $JS$ is written in terms of $S$. Therefore, generalizing $JS$ in Dempster-Shafer theory *entails* generalizing $S$ in the same theory. The hunt for a generalization of $S$ in Dempster-Shafer theory starts by noticing that *two* types of uncertainty coexist in this theory:

1) The *nonspecificity* in our prediction about the true world in a frame.
2) The *conflict* between the pieces of evidence expressed by each mass value.

To measure nonspecificity in Dempster-Shafer theory, we use generalized Hartley uncertainty functional [15], which is given in Definition 10.

*Definition 10: (Generalized Hartley Uncertainty Functional):* Let $m : \mathcal{P}(\mathcal{W}_s) \rightarrow [0,1]$ be a mass function on $\mathcal{W}_s$, and $\mathcal{F}_m$ be the set of all focal sets induced by $m$. The nonspecificity uncertainty about the true world in $\mathcal{W}_s$ is given by the functional:

$$GH(m) = \sum_{A \in \mathcal{F}_m} m(A) \log |A|$$

To *aggregately* measure both nonspecificity and conflict in Dempster-Shafer theory, we use the aggregate uncertainty functional [15], which is given in Definition 11.

*Definition 11 (Aggregate Uncertainty Functional):* Let $Bel : \mathcal{P}(\mathcal{W}_s) \rightarrow [0,1]$ be a belief function on $\mathcal{W}_s$. The aggregate uncertainty about the true world in $\mathcal{W}_s$ is given by the functional:

$$AU(Bel) = \max_{\mathcal{P}_{Bel}} \left\{ -\sum_{x \in \mathcal{W}_s} p(x) \log p(x) \right\}$$

where $\mathcal{P}_{Bel}$ is the set of all probability distribution functions that *dominate* $Bel$ by satisfying the following two properties:

1) $p(x) \in [0,1]$ for any $x \in \mathcal{W}_s$ and $\sum_{x \in \mathcal{W}_s} p(x) = 1$

2) $Bel(A) \leq \sum_{x \in A} p(x)$ for any $A \in \mathcal{P}(\mathcal{W}_s)$

A recursive algorithm for computing $AU$ is given in Appendix I-A [15]. It can be shown that $AU$ is *insensitive* to changes in evidence which makes it ill-suited for capturing the uncertainty associated with an agent's beliefs [15]. Therefore, $AU$ is not what we need in order to generalize $JS$ in Dempster-Shafer theory. However, If we recall that $AU$ is a total of two types of uncertainty; nonspecificity and conflict, we can write:

$$AU(Bel) = GH(m) + GS(m)$$

Based on this equivalence, we can define the generalized Shannon uncertainty functional.

*Definition 12: (Generalized Shannon Uncertainty Functional):* Let $m : \mathcal{P}(\mathcal{W}_s) \rightarrow [0,1]$ be a mass function, and $Bel : \mathcal{P}(\mathcal{W}_s) \rightarrow [0,1]$ be the corresponding belief function, both on $\mathcal{W}_s$. The conflict uncertainty about the true world in $\mathcal{W}_s$ is given by the functional:

$$GS(m) = AU(Bel) - GH(m)$$

where $GH(m)$ and $AU(Bel)$ are respectively given in Definitions 10 and 11.

Notice in Definition 12 that the insensitivity of $AU$ is *overcome* by subtracting $GH$ from it. This makes $GS$ sensitive to changes in evidence, and allows us to proceed with our *novel* generalization of $JS$ in Dempster-Shafer theory.

*Definition 13: (Generalized Jensen-Shannon Divergence Measure):* Let $m_1$ and $m_2$ be two mass functions on $\mathcal{W}_s$. The generalized Jensen-Shannon divergence measure between $m_1$ and $m_2$ is defined by the formula:

$$GJS(m_1, m_2) = 2GS(\frac{m_1 + m_2}{2}) - GS(m_1) - GS(m_2)$$

where $GS$ is given in Definition 12.

Now we have to check whether the properties of $JS$ listed in Table I hold on $GJS$. We know that for any $m$, we have $GS(m) \geq 0$, which means that $P1$ holds on $GJS$. $P2$ and $P3$ obviously hold on $GJS$. It is known that $GH(m) \leq \log |\mathcal{W}_s|$ and $AU(Bel) \leq \log |\mathcal{W}_s|$ for any $m$ and $Bel$ on $\mathcal{W}_s$ [15]. This means that $GS(m) \leq \log |\mathcal{W}_s|$ and consequently that $GJS(m_1, m_2) \leq \log |\mathcal{W}_s|$. Thus, $P4$ and $P6$ also hold.

V. LANGUAGE

We use an imperative while-language extended with a probabilistic choice construct. The language is described using rules that show how expressions and commands are formed, how expressions are evaluated, and how commands are executed.

*A. Syntax*

The syntactic sets and the metavariables that range over them are shown in Table II. The formation rules of arithmetic and Boolean expressions are standard, and we only give the formation rules of commands:

$c ::= \text{skip} | X := a | c_0; c_1 | \text{if } b \text{ then } c_0 \text{ else } c_1 | \text{while } b \text{ do } c | c_0 \,_p[] \, c_1$

The probabilistic choice rule $c_0 \,_p[] \, c_1$ executes $c_0$ with a probability $p$ or $c_1$ with a probability $1 - p$.

TABLE II
THE SYNTACTIC SETS AND THE METAVARIABLES

| Syntactic Set | Metavariables |
| --- | --- |
| $Val$: The set of integers $\mathbb{N}$ | $n,m$ |
| $Bool$: The set of truth values $\{\text{true}, \text{false}\}$ | $t$ |
| $Var$: The set of program variables | $X,Y$ |
| $Aexp$: The set of arithmetic expressions | $a$ |
| $Bexp$: The set of Boolean expressions | $b$ |
| $Com$: The set of commands | $c$ |

TABLE III
THE EXECUTION RULES OF COMMANDS

$[\text{skip}]\sigma \equiv \lambda \sigma \in State.\sigma$
$[X := a]\sigma \equiv \lambda \sigma \in State.\sigma[X \mapsto n]$ where $[a]\sigma = n$
$[c_0; c_1]\sigma \equiv ([c_1] \circ [c_0])\sigma = \lambda \sigma \in State.[c_1]([c_0]\sigma)$
$[\text{if } b \text{ then } c_0 \text{ else } c_1]\sigma \equiv \lambda \sigma \in State.([b]\sigma, [c_0]\sigma, [c_1]\sigma)$
$[\text{while } b \text{ do } c]\sigma \equiv \lambda \sigma \in State.$ least fixed point of $\Gamma : State \to State$ where $\Gamma(\varphi) = \lambda \sigma \in State.([b]\sigma, (\varphi \circ [c])\sigma, \sigma)$
$[c_0 \;_p[] \; c_1]\sigma \equiv \lambda \sigma \in State.p \times [c_0]\sigma + (1-p) \times [c_1]\sigma$

## B. Semantics

Recalling that a state in our scheme is an assignment of a value to a variable (what we mentioned in Section II-A), and having introduced the syntactic sets in the previous section, we can now denote a state as a function of the form $\sigma : Var \to Val$. When we write $\sigma(X) = n$ or $\sigma(X \to n)$ for $X \in Var$ and $n \in Val$, we mean that the value of the variable $X$ in the state $\sigma$ is $n$. We might have *more* than one variable in a single state, in which case we write $\sigma(X,Y) = (n,m)$ or $\sigma(X \to n, Y \to m)$ for $X,Y \in Var$ and $n,m \in Val$. A notation $State$ is also needed to refer to the set of all possible states in a program execution. We use the following semantic functions:

$$\mathcal{A} : Aexp \to (State \to Val)$$
$$\mathcal{B} : Bexp \to (State \to Bool)$$
$$\mathcal{C} : Com \to (State \to State)$$

which enables us to define the following denotation functions:

$$\forall a \in Aexp.\mathcal{A}[a] : State \to Val$$
$$\forall b \in Bexp.\mathcal{B}[b] : State \to Bool$$
$$\forall c \in Com.\mathcal{C}[c] : State \to State$$

Since the semantic functions are known, as well as the range of metavariables, we condense the denotations and write $[a]$, $[b]$, and $[c]$ instead of $\mathcal{A}[a]$, $\mathcal{B}[b]$, and $\mathcal{C}[c]$.

The evaluation of arithmetic and Boolean expressions is standard. As for commands, we note that their execution *changes* in program states. Unless the corresponding program inputs are influenced by an agent, we assume that variables in all states are initially set to *zero*, that is $\forall X \in Var.\sigma_0(X) = 0$. We also observe that a command execution may *terminate* in a final state, or may *diverge* and never yield a final state (non-termination). Let us explain the meaning of termination in this non-lifted semantics.

*Definition 14 (Non-lifted Meaning of Termination):* For any $c \in Com$, when we write:

$$[c]\sigma' = \lambda \sigma \in State.\sigma$$

we mean that the command $c$, which began in an input state $\sigma'$, *deterministically* terminates in an output state $\sigma$.

The execution rules of commands are given in Table III. The notion given in Definition 15 is used in one of those rules.

*Definition 15 (State Update):* Let $\sigma \in State$, $X \in Var$, and $n \in Val$ be a state, a variable, and a value respectively. The state obtained from $\sigma$ by changing the value of $X$ to $n$ in $\sigma$ is denoted as $\sigma[X \mapsto n]$. Formally, we write:

$$\sigma[X \mapsto n](Y) = \begin{cases} n & \text{if } Y = X \\ \sigma(Y) & \text{if } Y \neq X \end{cases}$$

We also make use of the simplifying and *colorful* notation:

$$(b, x, x') = \begin{cases} x & \text{if } b = \text{true} \\ x' & \text{if } b = \text{false} \end{cases}$$

## VI. LIFTED LANGUAGE

In this section, we lift the language we presented in Section V in order to act on mass functions. Our lifted language is the *first* of its kind to enjoy this property. The upgrade process involves both the syntax and the semantics.

### A. Lifting the Syntax

We need to add one more syntactic set to the sets shown in Table II, which is what we do in Definition 16.

*Definition 16 (The MASS Syntactic Set):* Let $\mathcal{W}_{h \cup l}$ be a frame on the overall variable set $h \cup l \subseteq r$ that contains a program's secret and nonsecret inputs. We define the syntactic set $MASS$ to be the set of all mass functions on $\mathcal{W}_{h \cup l}$, and we use the metavariables $m$ and $m'$ to range over $MASS$.

One more formation rule is also needed for any $m \in MASS$, and it is *luckily* prescribed in Definition 4.

### B. Lifting the Semantics

Assuming input (output) masses, when we write $m(\sigma) = n$ for $m \in MASS$, $\sigma \in State$ and $n \in [0,1]$, we mean the *likelihood* that $\sigma$ is to be used as an input (output) state. The only semantic function we need to lift is the one pertaining to commands. The lifted command semantic and denotation functions are defined by the mappings:

$$\mathcal{C} : Com \to (MASS \to MASS)$$
$$\forall c \in Com.\mathcal{C}[c] : MASS \to MASS$$

The meaning of termination also changes in the lifted semantics as shown in Definition 17.

*Definition 17 (Lifted Meaning of Termination):* For any $c \in Com$, when we write:

$$[c]m' = \lambda m \in MASS.\sum_\sigma m(\sigma).[c]\sigma$$

we mean that the command $c$, which began in *any* input state $\sigma'$ of $d(m')$, *potentially* terminates in any output state $\sigma$ of $d(m)$.

TABLE IV
THE LIFTED EXECUTION RULES OF COMMANDS

$[\text{skip}]m \equiv \lambda m \in MASS.m$
$[X := a]m \equiv \lambda m \in MASS.m[X \mapsto n]$ where $[a]\sigma = n$
for any $\sigma \in d(m)$
$[c_0; c_1]m \equiv ([c_1] \circ [c_0])m = \lambda m \in MASS.[c_1]([c_0]m)$
$[\text{if } b \text{ then } c_0 \text{ else } c_1]m \equiv \lambda m \in MASS.[c_0](m|b) + [c_1](m|\neg b)$
$[\text{while } b \text{ do } c]m \equiv \lambda m \in MASS.$ least fixed point of $\Gamma : MASS \to MASS$ where $\Gamma(\varphi) = \lambda m \in MASS.\varphi([c](m|b)) + (m|\neg b)$
$[c_0 \ _p[] \ c_1]m \equiv \lambda m \in MASS.[c_0](p \times m) + [c_1]((1-p) \times m)$

The sum value to the right-hand side of the previous formula specifies the likelihood of this termination.

In this context, we also need to give our *novel* definition of a mass update.

*Definition 18 (Mass Update):* Let $m \in MASS$, $X \in Var$, and $n \in Val$ be a mass function, a variable, and a value respectively. The mass function obtained from $m$ by changing the value of $X$ to $n$ in all the states of $d(m)$ is denoted as $m[X \mapsto n]$. Formally, we achieve that as follows:

1) $\forall \sigma \in d(m).\sigma' = \sigma[X \mapsto n] \in d(m[X \mapsto n])$
2) $m[X \mapsto n](\sigma') = \begin{cases} m(\sigma) & \text{if } X \in \sigma' \\ m(\sigma') & \text{if } X \notin \sigma' \end{cases}$

The lifted execution rules of commands are given in Table IV. These rules immediately follow from applying the formulas in definitions 17 and 18 to the execution rules given in Table III. Notice in the lifted rules that we are conditioning a mass function on a *Boolean expression*. Formula (3) can not do this. We give a *novel* adaptation of this formula in Definition 19.

*Definition 19 (Boolean Expression Conditioning):* Let $m : \mathcal{P}(\mathcal{W}_s) \to [0,1]$ be a mass function on $\mathcal{W}_s$, and $b \in Bexp$ be a Boolean expression. The expansion of $b$ to the domain $\mathcal{P}(\mathcal{W}_s)$ of $m$ yields the tuple set $B \subseteq \mathcal{W}_s$ whose tuples satisfy $b$ i.e., $B = \{x \in \mathcal{W}_s | x \vdash b\}$. The conditioning of $m$ on $b$ is then given by the formula:

$$m_b(A) = \begin{cases} \sum_{C \cap B = A} m(C) & \text{for } A \neq \emptyset \\ 0 & \text{for } A = \emptyset \end{cases}$$

Notice that the resulted mass function is *unnormalized*.

## VII. INFERENCE SCHEME

This sections presents an inference scheme an attacker uses to update her knowledge from interacting with a program execution. This scheme is a generalization of the experiment protocol advanced in [2]; however it *surpasses* that protocol by handling any number of secret inputs to a program. Before describing this scheme, we need to give the attacker's model.

### A. Attacker's Model

The attacker is modeled via the following assumptions:
1) The attacker has a *copy* of the program's source code.
2) The program has a *number* of secret inputs the attacker does not know and would like to learn.
3) The program executes on a system that does *not* intentionally collude to leak the secret inputs.
4) The program always terminates *and* preserves the state of secret inputs as high.
5) The program executes *once* per interaction with the attacker, and in each execution the attacker is allowed to make only *one* observation.
6) The attacker can monitor the public output of the program and *adaptively* change the input.
7) The attacker *knows* the frame of each secret input, *and* the values of all of the nonsecret inputs.
8) The impossible world is *not* a true value of any of the inputs [15]. Therefore, the attacker's belief is captured via a *normalized* mass function, which assigns a zero degree of belief in the impossible world (the empty set) as we saw in Definition 4.

### B. Scheme Description

At first, the attacker has an *initial* belief about the true values of the secret inputs. The extent of this belief is captured using an *initial* mass function $m_{init} : \mathcal{P}(\mathcal{W}_h) \to [0,1]$. This function can either reflect the attacker's initial *total ignorance* or her belief in an *initial* piece of evidence she obtained through some mean. In the former case, the attacker knows that the true values of the secret inputs are in the frame $\mathcal{W}_h$; however, she has no evidence whatsoever about their location in any subset of that frame, which gives $m_{init}(\mathcal{W}_h) = 1$ and $m_{init}(A) = 0$ for any $A \in \mathcal{P}(\mathcal{W}_h) \setminus \mathcal{W}_h$. In the latter case, the degree of the initial belief distributes (*unequally* in general) among a number of sets $I_1,...,I_m \in \mathcal{P}(\mathcal{W}_h)$ such that $m_{init}(I_1) = i_1 > 0,...,m_{init}(I_m) = i_m > 0$, $m_{init}(\mathcal{W}_h) = 1 - i_1 - ... - i_m$, and $i_1 + ... + i_m \leq 1$.

Without relying on monitoring a program execution, the attacker soon obtains a finite number $n$ of pieces of evidence (through social engineering say) from $n$ independent sources (independence was discussed in Section IV-A) about the true values of the secret inputs. The extent of these $n$ pieces of evidence is captured using $n$ mass functions $m_i : \mathcal{P}(\mathcal{W}_h) \to [0,1]$ where $i = 1,...,n$.

Before experimenting with a program execution, the attacker ought to combine the mass functions she has using formula (1). The combination outcome is the attacker's *prebelief* $m_{pre}$, which describes her belief *before* interacting with the program:

$$m_{pre} : \mathcal{P}(\mathcal{W}_h) \to [0,1] : m_{pre}(A) = m_{init} \otimes \bigotimes_{i=1}^{n} m_i(A)$$

The system chooses the *high* projection of the input state $\sigma^{\downarrow h} \in \mathcal{P}(\mathcal{W}_h)$ to be the set that contains the true values of the secret inputs. The corresponding point mass function would be:

$\dot{m}_h : \mathcal{P}(\mathcal{W}_h) \to [0,1] : \dot{m}_h(\sigma^{\downarrow h}) = 1, \dot{m}_h(A) = 0$ for any $A \in \mathcal{P}(\mathcal{W}_h) \setminus \sigma^{\downarrow h}$

The attacker chooses the *low* projection of the input state $\sigma^{\downarrow l} \in \mathcal{P}(\mathcal{W}_l)$ with the corresponding point mass function:

$\dot{m}_l : \mathcal{P}(\mathcal{W}_l) \to [0,1] : \dot{m}_l(\sigma^{\downarrow l}) = 1, \dot{m}_l(A) = 0$ for any $A \in \mathcal{P}(\mathcal{W}_l) \setminus \sigma^{\downarrow l}$

The low projection $\sigma^{\downarrow l}$ represents the attacker's *guesses* of the secret inputs, in addition to the nonsecret inputs. These guesses are likely to be *influenced* by the attacker's prebelief, in which case, the attacker would choose $\sigma^{\downarrow l}$ as the set that has the highest mass according to $m_{pre}$. However, we do not impose an influence as such to avoid the loss of generality.

The program's input becomes the combination $\dot{m}_h \otimes \dot{m}_l$ done using formula (2), since the domains of $\dot{m}_h$ and $\dot{m}_l$ are different. The system executes the program which produces a mass function:

$$m_\delta : \mathcal{P}(\mathcal{W}_{h \cup l}) \to [0,1] : m_\delta(A) = [S](\dot{m}_h \otimes \dot{m}_l)(A)$$

This mass function represents *many* possible output states. However, since the attacker is allowed to make only one observation per execution, one state must be chosen *randomly*. This random choice is made using a sampling operator $\Gamma$ that draws a state $\sigma'$ from the domain of $m_\delta$ with a probability $1/|\mathcal{F}_{m_\delta}|$. The chosen output state becomes $\sigma' \in \Gamma(m_\delta)$, from which the attacker *observes* the low projection $o = \sigma'^{\downarrow l} \in \mathcal{P}(\mathcal{W}_l)$.

The attacker applies the semantics of the program to the combination $\dot{m}_l \otimes m_{pre}$ to generate a *prediction* $m'_\delta$ of the output mass $m_\delta$:

$$m'_\delta : \mathcal{P}(\mathcal{W}_{h \cup l}) \to [0,1] : m'_\delta(A) = [S](\dot{m}_l \otimes m_{pre})(A)$$

The attacker incorporates any additional information contained in the observation $o$, she made earlier, by conditioning $m'_\delta$ on $o$ using formula (3). The result is a new mass function $m''_\delta$ the attacker projects to $h$ to obtain her *postbelief* $m_{post} = m''^{\downarrow h}_\delta$, which describes her belief after interacting with the program.

It is worth pointing out that in repeated executions, the attacker may choose her *postbelief* from one execution as a prebelief to the next. The attacker may even choose a prebelief that *contradicts* the pieces of evidence she has. Both choices are acceptable and add ample expressiveness to our analysis.

## VIII. Experimenting with the Inference Scheme

Unlike the QIF method used in [2], which can only deal with singleton focal sets induced by an attacker's beliefs, our method is capable of handling all focal set structures. This includes, in addition to singleton focal sets, focal sets that form a *partition*, *overlapping*, and *nested* focal sets. Experimenting with our scheme using singleton sets yields identical results to those in [2]. We also find it rather similar to experiment using overlapping or nested sets. Therefore, we experiment with only partition and nested sets. For the purpose of our experiments, we reuse the same password checker from [2]. This checker sets an authentication flag $a$ after checking a stored password $p$ against a guessed password $g$ supplied by the user.

$$\mathcal{PWC} : \text{if } p = g \text{ then } a := 1 \text{ else } a := 0$$

The secret input to this $\mathcal{PWC}$ is $p$ while the nonsecret ones are $g$ and $a$. The universal variable set is $r = \{p,g,a\}$ and the high and low variable sets are $h = \{p\}$ and $l = \{g,a\}$ respectively. For simplicity, $p$ is assumed to be either $A$, $B$, or $C$. Each conducted experiment involves two runs of interaction between the attacker and $\mathcal{PWC}$. The *real* password is assumed to be $A$ in the first run and $C$ in the second.

TABLE V
THE ATTACKER'S PREBELIEF AND POSTBELIEF IN EXPERIMENT 1

| $\mathcal{P}(\mathcal{W}_h)$ | $m_{pre}$ | $m'_{post}$ | $m''_{post}$ |
|---|---|---|---|
| $\{A\}$ | .98 | 1 | 0 |
| $\{B,C\}$ | .02 | 0 | 1 |

TABLE VI
AN INTERMEDIATE TABLE FOR COMPUTING $\dot{m}_h \otimes \dot{m}_l$

|  | $\{(A,A,0),(B,A,0),(C,A,0)\} : 1$ |
|---|---|
| $\{(A,A,0),(A,A,1),(A,B,0),$ $(A,B,1),(A,C,0),(A,C,1)\} : 1$ | $\{(A,A,0)\} : 1$ |

### A. Experiment 1

In this experiment, the focal sets induced by $m_{pre}$ form a partition as shown in Table V. Notice that the attacker believes $p$ is *overwhelmingly* likely to be $A$, but has a very small chance (*not* necessarily equally distributed) to be either $B$ or $C$.

*1) Interaction 1:* The system chooses $\sigma^{\downarrow h} = (p \to A)$ and the attacker chooses $\sigma^{\downarrow l} = (g \to A, a \to 0)$. The corresponding $\dot{m}_h$ and $\dot{m}_l$ are given in Section VII-B. The program input $\dot{m}_h \otimes \dot{m}_l$ is determined by applying formula (2). We simplify this task by performing the intermediate computations shown in Table VI [12]. The first column in this table contains $\dot{m}_h$ and the top row contains $\dot{m}_l$, both of which *extended* to the union variable domain $h \cup l = \{p,g,a\}$. Every internal cell contains the *intersection* between the corresponding tuple sets and the *product* of the corresponding values. The combination is finalized by adding the values of all internal cells with equal tuple set and normalizing by $k = 1$ to obtain:

$$\dot{m}_h \otimes \dot{m}_l = [\{(A,A,0)\} : 1]$$

Next the semantics of $\mathcal{PWC}$, given in Table IV, is applied:

$$[\mathcal{PWC}](\dot{m}_h \otimes \dot{m}_l) = [c_0]((\dot{m}_h \otimes \dot{m}_l)|b) + [c_1]((\dot{m}_h \otimes \dot{m}_l)|\neg b)$$

where:

$$c_0 ::= a := 1, c_1 ::= a := 0, b ::= p = g, \neg b ::= p \neq g \quad (4)$$

The expansion of $b$ to $\mathcal{P}(\mathcal{W}_{h \cup l})$ yields:

$$B = \{(A,A,0),(A,A,1),(B,B,0), \\ (B,B,1),(C,C,0),(C,C,1)\} \quad (5)$$

Applying Definition 19 conditioning gives:

$$(\dot{m}_h \otimes \dot{m}_l)|b = [\{(A,A,0)\} : 1]$$

The expansion of $\neg b$ to $\mathcal{P}(\mathcal{W}_{h \cup l})$ yields:

$$\neg B = \{(A,B,0),(A,B,1),(A,C,0),(A,C,1), \\ (B,A,0),(B,A,1),(C,A,0),(C,A,1), \quad (6) \\ (B,C,0),(B,C,1),(C,B,0),(C,B,1)\}$$

Applying Definition 19 conditioning again gives:

$$(\dot{m}_h \otimes \dot{m}_l)|\neg b = [\emptyset : 1]$$

Now we apply the mass updates, as described in Definition 18:

$$[c_0]((\dot{m}_h \otimes \dot{m}_l)|b) = [\{(A, A, 1)\} : 1]$$
$$[c_1]((\dot{m}_h \otimes \dot{m}_l)|\neg b) = [\emptyset : 1]$$

A straightforward addition gives:

$$[\mathcal{PWC}](\dot{m}_h \otimes \dot{m}_l) = [\{(A, A, 1)\} : 1; \emptyset : 1]$$

and a final normalization yields the output mass:

$$m_\delta = [\mathcal{PWC}](\dot{m}_h \otimes \dot{m}_l) = [\{(A, A, 1)\} : 1]$$

The only state that can be drawn from $d(m_\delta)$ is $(A, A, 1)$ from which the attacker observes the low projection:

$$\sigma' = (p \to A, g \to A, a \to 1)$$
$$o = \sigma'^{\downarrow l} = (g \to A, a \to 1)$$

Next $\dot{m}_l \otimes m_{pre}$ is determined by applying formula (2):

$$\dot{m}_l \otimes m_{pre} = [\{(A, A, 0)\} : .98; \{(B, A, 0), (C, A, 0)\} : .02]$$

The semantics of $\mathcal{PWC}$ is now applied to get:

$$[\mathcal{PWC}](\dot{m}_l \otimes m_{pre}) = [c_0]((\dot{m}_l \otimes m_{pre})|b) + [c_1]((\dot{m}_l \otimes m_{pre})|\neg b)$$

where $c_0$, $c_1$, $b$, and $\neg b$ are the same as in (4). Applying Definition 19 conditioning with the same (5) and (6) yields:

$$(\dot{m}_l \otimes m_{pre})|b = [\{(A, A, 0)\} : .98; \emptyset : .02]$$
$$(\dot{m}_l \otimes m_{pre})|\neg b = [\{(B, A, 0), (C, A, 0)\} : .02; \emptyset : .98]$$

The mass updates are now applied to get:

$$[c_0]((\dot{m}_l \otimes m_{pre})|b) = [\{(A, A, 1)\} : .98; \emptyset : .02]$$
$$[c_1]((\dot{m}_l \otimes m_{pre})|\neg b) = [\{(B, A, 0), (C, A, 0)\} : .02; \emptyset : .98]$$

A straightforward addition gives:

$$[\mathcal{PWC}](\dot{m}_l \otimes m_{pre}) = [\{(A, A, 1)\} : .98;$$
$$\{(B, A, 0), (C, A, 0)\} : .02; \emptyset : 1]$$

and a final normalization yields the attacker's prediction:

$$m'_\delta = [\mathcal{PWC}](\dot{m}_l \otimes m_{pre}) = [\{(A, A, 1)\} : .98;$$
$$\{(B, A, 0), (C, A, 0)\} : .02]$$

After expanding $o$ to $d(m'_\delta)$ and obtaining:

$$O = \{(A, A, 1), (B, A, 1), (C, A, 1)\}$$

the attacker conditions using formula (3) to get:

$$m''_\delta = m'_\delta | o = [\{(A, A, 1)\} : 1] \text{ where } k = 1/.98$$

A final projection of $m''_\delta$ to $h$ yields $m'_{post}$ shown in Table V.

*2) Interaction 2:* Similar computations to those presented in the previous section yields $m''_{post}$, also shown in Table V.

TABLE VII
THE ATTACKER'S PREBELIEF AND POSTBELIEF IN EXPERIMENT 2

| $\mathcal{P}(\mathcal{W}_h)$ | $m_{pre}$ | $m'_{post}$ | $m''_{post}$ |
|---|---|---|---|
| $\{A, B\}$ | .98 | 0 | 0 |
| $\{A, B, C\}$ | .02 | 0 | 0 |
| $\{A\}$ | 0 | 1 | 0 |
| $\{B\}$ | 0 | 0 | .98 |
| $\{B, C\}$ | 0 | 0 | .02 |

*3) Reasoning About the Results:* If we contemplate the results in Table V. $m'_{post}$ suggests that the attacker is certain that $p$ is $A$, whereas $m''_{post}$, suggests that she is certain that the $p$ is either $B$ or $C$ (with chances that are not necessarily equal). Comparing $m'_{post}$ with $m_{pre}$ tells that interaction 1 had begotten little change in the attacker's belief. This little *change* corresponds to little *update* in the attacker's knowledge and subsequently to little information flow from $\mathcal{PWC}$. If we compare $m''_{post}$ with $m_{pre}$, we arrive at the converse conclusion - larger knowledge update and larger flow. Notice also that $m'_{post}$ and $m''_{post}$ are more accurate than $m_{pre}$ since both of them are *nearer* to $\dot{m}_h$ than it. This accuracy increase results in informing of the attacker, which is *positive* information flow.

*B. Experiment 2*

In this experiment, the focal sets induced by $m_{pre}$ are nested as shown in Table VII. Notice that the attacker believes $p$ is overwhelmingly likely to be either $A$ or $B$, but has a very small chance to be either $A$, $B$, or $C$ (all the chances are not necessarily equal). The attacker's postbeliefs $m'_{post}$ and $m''_{post}$ are shown in the same table.

*1) Reasoning About the Results:* If we contemplate the results in Table VII. $m'_{post}$ suggests that the attacker is certain that $p$ is $A$, whereas $m''_{post}$ suggests that she believes $p$ is overwhelmingly likely to be $B$ but has a very small chance (not necessarily equally distributed) to be either $B$ or $C$. Comparing $m'_{post}$ with $m_{pre}$ tells that interaction 1 had begotten large change in the attacker's belief (she no longer believes in $\{B\}$). This large change corresponds to large knowledge update and large flow. Comparing $m''_{post}$ with $m_{pre}$ yields the converse conclusion. Notice also that $m'_{post}$ is more accurate than $m_{pre}$ since it is nearer to $\dot{m}_h$ than it. This accuracy increase results in informing of the attacker, and means positive information flow. However, we *cannot* informally claim that $m''_{post}$ is more accurate than $m_{pre}$ - they both seem to stand at nearly the same distance from $\dot{m}_h$ (which is a point mass on $\{C\}$). This *nearly-constant* accuracy reflects *near-zero* information flow.

*C. Generic Observations*

We can derive generic *and* informal observations by putting the experiments' results into a wider perspective. If the attacker has a *strong* belief that the true value of a secret input is in a partition (in a set nested in other sets in the body of evidence), and an interaction with the system *confutes* her belief, then the attacker's strong belief is transferred to that partition's *complement* (those sets' *intersection*).

## IX. MEASURING INFORMATION FLOW

The approach used in [2] to measure information flow, which corresponds it to an improvement in the accuracy of an attacker's belief, is applicable in our setting. Recall from Section IV-C2 that $GJS(m_1, m_2)$ measures the divergence between $m_1$ and $m_2$. The accuracy of the attacker's prebelief $m_{pre}$ is its distance from $\dot{m}_h$, measured as $GJS(m_{pre}, \dot{m}_h)$. Likewise, the accuracy of the attacker's postbelief $m_{post}$ is $GJS(m_{post}, \dot{m}_h)$. We define the amount of information flow $\mathcal{Q}$ as the difference between these two quantities:

$$\mathcal{Q} = GJS(m_{pre}, \dot{m}_h) - GJS(m_{post}, \dot{m}_h)$$
$$= 2GS(\frac{m_{pre}+\dot{m}_h}{2}) - 2GS(\frac{m_{post}+\dot{m}_h}{2})$$
$$- GS(m_{pre}) + GS(m_{post})$$

Calculating the amount of flow from the experiments conducted in Section VIII yields .020145, .97999, 1.01999, and .01999 bits respectively [10]. These results are in line with the informal reasoning made in sections VIII-A3 and VIII-B1.

Unlike the metric proposed in [2], our measure has an intrinsic *absolute* range bounded by the size of a program's secret input as proved in Theorem 1.

*Theorem 1:* Considering both deterministic and probabilistic programs, and all types of an attacker's beliefs, and *avoiding* the imposition of any admissibility restriction on those beliefs, the general range of flow reported by $\mathcal{Q}$ is:

$$\varrho_\mathcal{Q} = [-\eta, \eta]$$

where $\eta$ is the size of a program's secret input in bits.

*Proof:* The proof is given in Appendix I-B. ∎

Additionally, the results reported by our measure are easily associated with the exhaustive search effort needed to uncover a program's secret information. This can be easily shown by assuming a program with a secret input of size $\eta$ bits, and an informing flow of $k$ bits from the same program to an attacker. The absolute upper bound of $\mathcal{Q}$, given in Theorem 1, tells us that $k \leq \eta$. Therefore, the space of the exhaustive search [16] that should be carried out in order to reveal the residual part $\eta - k$ bits of the secret input is $2^{\eta-k}$. On the contrary, our earlier work [7] showed that the exhaustive search space cannot be established under the metric proposed in [2].

## X. CONCLUSIONS

We presented a generalization of the QIF analysis method proposed in [1], [2]. Our generalization is based on Dempster-Shafer theory of imprecise probabilities. We uncovered a number of weaknesses in the original method and showed that they are eliminated by way of our generalization. Our generalized method can handle any number of secret inputs to a program, it enables the capturing of an attacker's beliefs in all kinds of sets (singleton or not), and it supports a new and precise QIF measure whose reported flow results are plausible in that they are bounded by the size of a program's secret input, and can be easily associated with the exhaustive search effort needed to uncover a program's secret information, unlike the results reported by the original metric.

## ACKNOWLEDGMENT
The author would like to thank Marc Pouly for his helpful comments on an early draft of this paper.

## APPENDIX I
## ALGORITHMS AND PROOFS

### A. Computing Aggregate Uncertainty

*Input*: A belief function $Bel : \mathcal{P}(\mathcal{W}_s) \to [0, 1]$ on $\mathcal{W}_s$.
*Output*: $AU(Bel)$ as given in Definition 11.

1) Find a nonempty set $A \in \mathcal{P}(\mathcal{W}_s)$ such that $Bel(A)/|A|$ is maximal. If more than one set exist, assume the set that has the highest cardinality.
2) For any $x \in A$, put $p(x) = Bel(A)/|A|$.
3) For each $B \subseteq \mathcal{W}_s - A$, put $Bel(B) = Bel(B \cup A) - Bel(A)$.
4) Put $\mathcal{W}_s = \mathcal{W}_s - A$.
5) If $\mathcal{W}_s \neq \emptyset$ and $Bel(\mathcal{W}_s) > 0$, go to step 1.
6) If $\mathcal{W}_s \neq \emptyset$ and $Bel(\mathcal{W}_s) = 0$, put $p(x) = 0$ for any $x \in \mathcal{W}_s$.
7) Compute $AU(Bel) = -\sum_{x \in \mathcal{W}_s} p(x) \log p(x)$.

### B. Proof of Theorem 1

$$0 \leq GJS(m_1, m_2) \leq \log |\mathcal{W}_s| = \eta \text{ (from Section IV-C2)}$$
$$-\eta \leq GJS(m_{pre}, \dot{m}_h) - GJS(m_{post}, \dot{m}_h) \leq \eta$$
$$\varrho_\mathcal{Q} = [-\eta, \eta]$$